\documentclass[referee]{aa}
\usepackage{graphicx}
\usepackage{txfonts}
\usepackage{natbib}

\begin{document}

  \title{A multi--wavelength study of the IRAS Deep Survey galaxy sample}

   \subtitle{II. The far-IR properties.}  

   \author{P. Mazzei
  \inst{1} 
  \and
   A. della Valle\inst{1,2}
   \and
     D. Bettoni\inst{1}
           }

    \offprints{P. Mazzei}

    \institute{INAF, Padova Astronomical Observatory, Vicolo dell'Osservatorio, 5
    Padova, 35128, Italy\\
         \email{paola.mazzei,antonio.dellavalle,daniela.bettoni@oapd.inaf.it}
         \and
           Department of Astronomy , Padova, Italy }

          \date{}

\abstract 	
{The luminosity function (LF) is a basic tool in the study of galaxy evolution since it
constrains galaxy formation models.
The earliest LF estimates in the IR and far-IR spectral ranges seem to suggest 
strong evolution. Deeper samples are needed to confirm these predictions.
We have a useful  IR  dataset, which provides a
direct link between IRAS and ISO surveys, and the forthcoming deeper
Spitzer Space Telescope and  Akari cosmological surveys, to
address this issue.
}
{
This allows us to derive the 60\,$\mu$m local LF to sensitivity 
levels 10 times deeper than before, to investigate evolutionary effects up to 
a redshift of 0.37,  and,
using the  60/15$\mu$m bi-variate method,  the poorly known 15\,$\mu$m local LF
of galaxies.
 }
{ 
We exploited our ISOCAM observations of the IRAS Deep Survey (IDS) fields (Hacking \& Houck
1987), 
to correct the $60\,\mu$m fluxes for confusion effects and
observational biases.  We find indications of a
significant incompleteness of the IDS sample, 
still one of the deepest far-IR selected galaxy samples,
below $\simeq 80\,$mJy (Mazzei et al. 2001).  
We have reliable identifications and spectroscopic redshifts for
100\% of  a complete subsample comprising 56 sources with
$S(60\mu\hbox{m})> 80\,$mJy.
}{ With our spectroscopic coverage we construct the 60\,$\mu$m LF for a sample
complete down to 80 mJy. This LF extends
over three orders of magnitude in luminosity, from 9 up to more than 12
in $\log(L_{60}/L_{\odot})$.
Despite the fact that the redshift range of our sample exceeds $z=0.3$, 
the $V/V_{max}$ test  gives $<V/V_{max}>=0.51\pm 0.06$, consistent with 
a uniform distribution of sources. 
A more direct test, whereby the LF was measured
in each of four different redshift intervals, does not point out any  signature of 
evolution. On the other hand, the
rest--frame 15$\mu$m local LF we derive, extends up to $\log(L_{15}/L_{\odot})=12$ and predicts
10 times more sources at $\log(L_{15}/L_{\odot})=11$  
than are seen by  Pozzi et al. (2004).
}
{}

	
 	\authorrunning{Mazzei et al.}
 
	\titlerunning{Multi--wavelength study of the IDS/ISOCAM sample.II}

 \maketitle
%


\section{Introduction}\label{intro}

Understanding how  galaxies form and evolve is a key goal of  modern 
physical cosmology.  A fundamental observable of galaxies is their luminosity function 
(LF) which  has long been used to constrain galaxy formation models and to quantify
 star formation  and evolution both in luminosity and in density.
 The IR and far-IR  spectral ranges are the best to deepen our knowledge  of
this subject since  they trace the 
star formation that is responsible for galaxy formation.
In particular several satellite missions, in
the past (IRAS and ISO), and in the present (Spitzer, Akari), 
provided and will provide  data  which will be complementary 
for a  detailed study of the LFs in such  spectral domains.

The earliest IR estimate of the  LFs, derived from the IRAS data \citep{RRW87,
Sau00}, indicated   strong evolution, so that LF increases with redshift
 $\propto (1+z)^{3 \pm 1}$. Moreover
deep surveys at 15$\mu$m carried out with ISO  
\citep[i.e.,][]{Elbazetal99, Flores99, Larietal01, Metcalfe2003} 
seem to require, indeed,  strong
evolution of 15$\mu$m sources 
(Lagache et al. (2005) and references therein). Several
evolutionary models were developed to explain these results,
\citep[e.g.,][]{Fetal01,RR01} and fit the IR/submillimiter source counts with different
degrees of success. Nevertheless none of them is based on a local LF obtained 
from 15$\mu$m data, since the only available data until  recently came from
IRAS 12$\mu$m photometry \citep{Rush93, xu98, Fang98}. 
The first attempt to build up the 15$\mu$m LF of a NEPR subsample was made by
Xu (2000). However  he said 
that it must be considered as a preliminary work because:
i) the sample of galaxies used is a incomplete sample, ii) there is a 
possible misidentification between the sources in the 60$\mu$m redshift 
survey of Ashby et al. (1996) and the 15$\mu$m sources in his work  (see della Valle et al. (2006), for more
details), iii) the model 
used to interpret the data treated all IR galaxies as a single population. 

Our IDS/ISOCAM sample overcomes all these issues.
It  comprises a complete, 60$\mu$m  selected sample of 56 galaxies in the North 
Ecliptic Polar Region (NEPR),
a subsample of the original  98 IRAS Deep Survey (IDS) fields \cite{HH87}.

The IRAS Deep Survey (IDS) sample was defined by co-adding IRAS
scans of the North Ecliptic Polar Region (NEPR), representing more
than 20 hours of integration time \cite{HH87}.  It comprises 98
sources with $S(60\mu\hbox{m})> 50\,$mJy over an area of 6.25
square degrees. 
 
Mazzei et al. (2001) exploited  ISOCAM observations 
(range 12-18$\mu$m) of 94 IRAS Deep Survey (IDS) fields \citep{Aus99}, 
centered on the nominal positions of IDS sources, 
to correct the $60\,\mu$m fluxes for confusion effects and
observational biases,   finding indications of a
significant incompleteness of the IDS sample
below $\simeq 80\,$mJy. 
In della Valle et al (2006) we presented spectroscopic and
optical observations of candidate identifications  of our
ISOCAM sources. 
Combining such observations with those by Ashby et al. (1996),
we have reliable identifications and spectroscopic redshifts for
100\% of the complete subsample comprising 56 sources with
$S(60\mu\hbox{m})> 80\,$mJy. 
It is 
the deepest complete IRAS selected sample available and still one
of the deepest complete far-IR selected samples. For
comparison, the IRAS {\it Point Source Catalog} 
\citep[hereafter PSC;][]{Beichman1988}, 
comprises about $250,000$ IR sources with a completeness
limit of 0.5 Jy at 60$\mu$m  \citep{Soi87}.
The deep ISOPHOT surveys, FIRBACK at 170$\mu$m,
(Puget et al. 1999; Dole et al. 2001), and ELAIS at 90$\mu$m 
\citep{Oliver00} are all complete down to about 100\,mJy. Moreover,
the $70\,\mu$m Spitzer catalog of the 8.75 sq. deg. Bootes field
is flux limited to 80 mJy (Dole et al. 2004) and the Spitzer
extragalactic ``main" First Look Survey, covering about 4 sq.
deg., is complete to about 20 mJy at $70\,\mu$m (cf. Fig. 2 of
Frayer et al. 2006), but redshift measurements are available for a
substantial fraction of sources (yet only 72\%) merely for
$S_{70\mu{\rm m}} > 50\,$mJy.

Thanks to our ISOCAM and optical/near-IR
observations  our sample, which provides a
direct link between IRAS and ISO surveys, and the forthcoming deeper
Spitzer Space Telescope and  Akari cosmological surveys 
\footnote{New deep observations of the NEPR are planned with the Akari 
space mission, also known as the InfraRed Imaging
Surveyor (IRIS). It will map the entire sky in four far-IR bands,
from 50 to 200$\mu$m, and two mid-IR bands, at 9 and 20$\mu$m,
with far-IR angular resolutions of 25--45 arcsec, reaching a
detection limit of 44 mJy ($5 \sigma$ sensitivity) in the
50--75$\mu$m band \citep{Pearson2004}. With the Akari orbit, the
integration time on the NEPR will be particularly high, and,
correspondingly, the detection limit significantly deeper than
average.}, 
 is one of the far-IR selected complete samples 
with the larger spectral coverage. 
In addition to the ISOCAM and to 
the 60$\mu$m fluxes, most of them $\simeq$70\%   have 100$\mu$m fluxes,
the remaining 30\% with upper limits,
and several ($\simeq$40\%) have 25$\mu$m fluxes from IRAS \cite{Maetal01}. 
Optical imaging has been already performed for 62.5\% out 
of such a sample  in at least one band, B or R, and for 34\% 
in both the bands,  moreover 
Two Micron All Sky Survey (2MASS)
data are available for 68\% of the sample and 
VLA observations for a large fraction of these
sources are also available \citep{H89}. 

In this paper, which is the  second step of  our multi-wavelength approach devoted to
study  the evolution of a far-IR selected sample of galaxies on which numerous studies of the 
far-IR evolution of galaxies still rely,
we derive  the 60$\mu$m luminosity function (LF) of such
a sample. Our sample, ten times deeper in flux density than the PSC catalog,
thus less liable to the effect of local density inhomogeneity, allows us to investigate 
evolutionary effects up to a redshift of 0.37,  five times deeper than the
PSCz catalog  \cite[z$\le$ 0.07,][]{Sau00}. 
Moreover, we use the bi-variate method
to translate the 60$\mu$m  LF to the poorly known  15$\mu$m LF.
We will compare our results with the  recent determination of the 15$\mu$m local LF 
obtained by  Pozzi et al. (2004)  using the available data on the 
southern  fields, S1 and S2, of the ELAIS survey \cite{Oliver00}.

The plan of the paper is the following: Section 2 focuses on far-IR properties
of our complete sample, Section 3
shows our derived 60\,$\mu$m LF, Section 4 presents the 15\,$\mu$m LF based on
the bi-variate 60/15\,$\mu$m method.
In Section 5 there are our conclusions.\\
Here and in the following we adopt: 
$\Lambda$=0.7, $\Omega_{b}$=0.3, $H_0$=70 km/s/Mpc.

\section{The far-IR properties}\label{farir}
Our complete sample 
comprises 56 IDS/ISOCAM sources. 
25\%  out of  these
are beyond $z=0.1$, 12.5\%   beyond $z=0.15$ and only 5.3\%  at $z>0.2$  
(Paper I). Such a sample is deeper than previous estimates 
\cite{ash96},
showing a tail extending up to z=0.375, almost  4 Gyr in
look--back time. 

Our morphological analysis \citep{bettoni05} 
shows that,  although   16\% of our sources are multiple systems,
unperturbed disk galaxies dominate the IDS/ISOCAM sample. 
One ULIRG, 3-53A,
and two broad H$\alpha$ emission line galaxies with AGN optical properties
(i.e., 3-70A and 3-96A, see Bettoni et al. (2006) for more details), are also 
included in the complete subsample.

\begin{figure*}
\centering
\includegraphics[width=.48\textwidth]{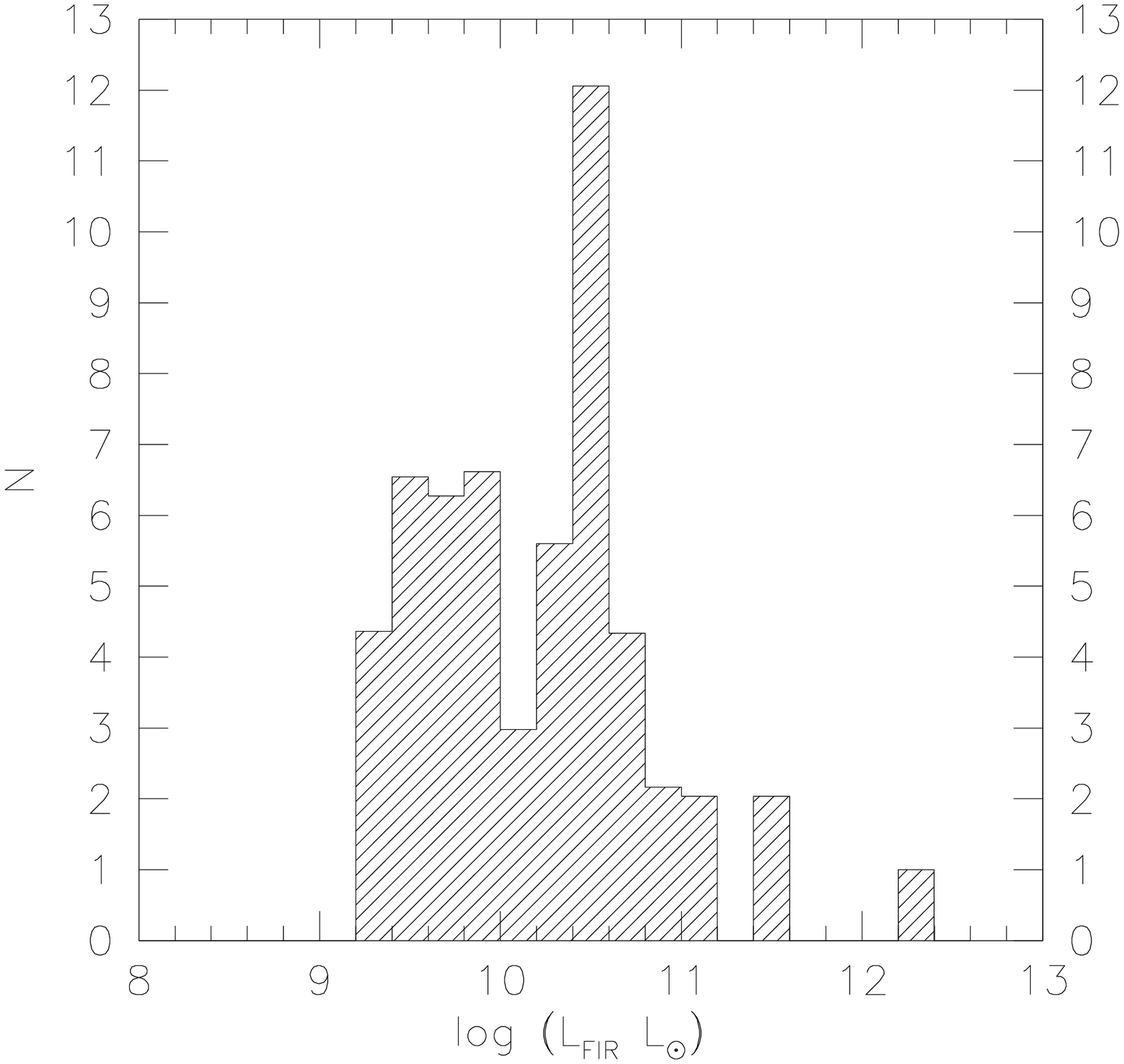}
\includegraphics[width=.48\textwidth]{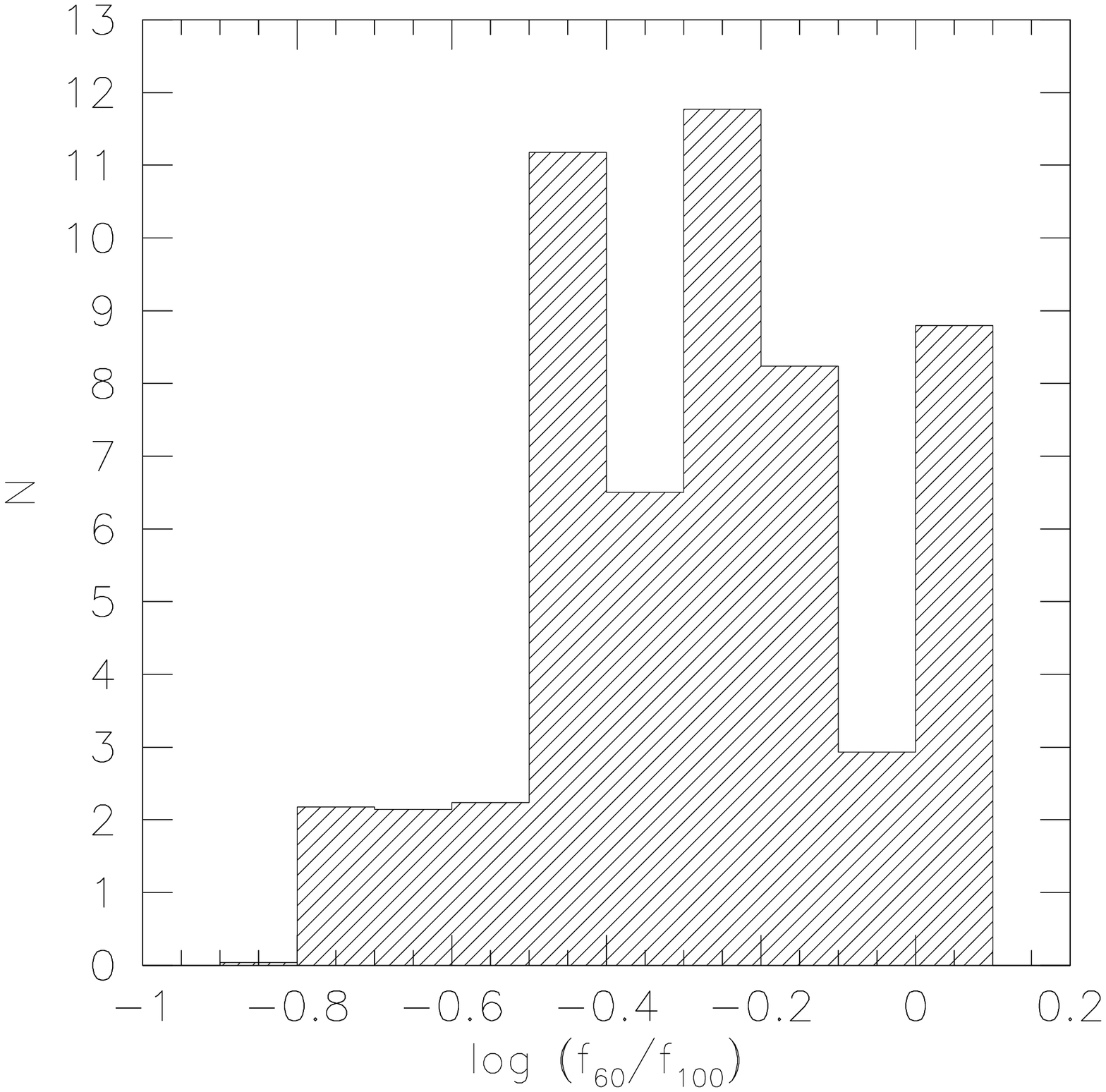}
\caption[]{
{\sl Left:} The rest--frame far-IR luminosity distribution of the IDS/ISOCAM sample of 56 
galaxies.  Here as in Figures \ref{Lfir}, \ref{f60100}, \ref{eps4}, 
\ref{pearson}, \ref{pearson_sbs} and \ref{lf15_sbs}, upper limits are taken into
account by exploiting the Kaplan-Mayer estimator and by accounting for
K-corrections and dust emission  using evolutionary population
synthesis models (see text). 
{\sl Right:} The rest--frame distribution of the IRAS flux density ratio 
f$_{60}/$f$_{100}$, as revised in Mazzei et al (2001), for the same sample.}
\label{eps1}
\end{figure*}

Fig. \ref{eps1} (left panel) shows the distribution of the far-IR luminosity 
($L_{FIR}$, from
42.5 to 122.5 \,$\mu$m)  of  
our sample,  where  $L_{FIR}=4\pi D_L^2$(FIR), 
FIR=1.26$\times 10^{-14}(2.58f_{60}+f_{100})$\,W/m$^2$, and $f_{60}$ and $f_{100}$
are in Jy \cite{he88}.
In such a figure, 
as in the following ones, 
K-corrections were derived from  evolutionary population synthesis models taking into account
dust effects \cite{Maz95}, luminosities were in units of solar bolometric 
luminosity, L$_{\odot}=3.83\times 10^{33}$\,erg/s. Moreover, 
upper limits to flux densities were
accounted for  by exploiting the Kaplan-Meier estimator \cite{KM58}.
Calculations were carried out using the ASURV v 1.2 package
\cite{Isobe} 
which implements methods presented in Feigelson \& Nelson (1985) and in
Isobe et al. (1986). 
The Kaplan-Meier estimator is a non-parametric, 
maximum-likelihood-type estimator of the ``true'' distribution function 
(i.e., with all quantities properly measured,  and no upper limits). The 
``survivor'' function, giving the estimated proportion of objects with upper
limits falling in each bin, does not produce, in general, integer numbers,
but is normalized to the total number.  This is why non-integer numbers of objects
appear in the histograms of our figures.

\begin{figure*}
\centering
\includegraphics[width=.48\textwidth]{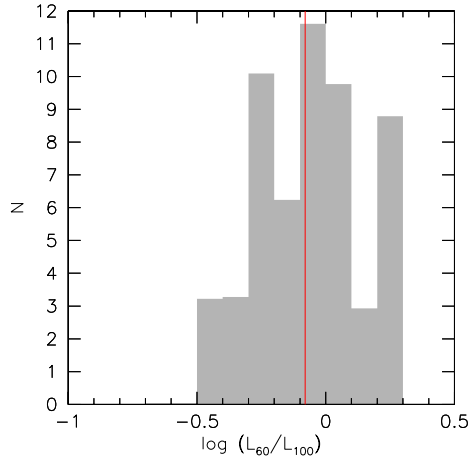}
\caption[]{
The distribution of the rest--frame luminosity ratio, $\log L_{60}/L_{100}$, for our
56 IDS/ISOCAM sources; red continuous line corresponds to the value $\log
L_{60}/L_{100}=-0.08$ (see text).
}
\label{Lfir}
\end{figure*}

\begin{figure*}
\centering
\includegraphics[width=.45\textwidth]{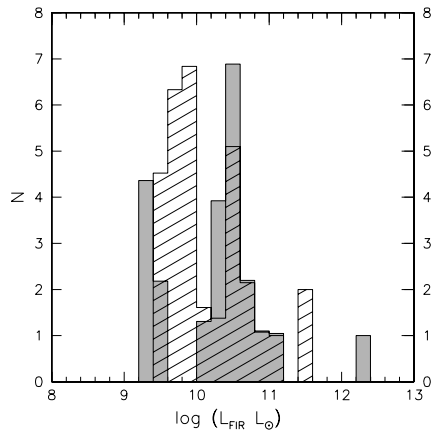}
\caption[]{
The rest-frame far-infrared luminosity distribution for the IDS/ISOCAM sample of
56 galaxies.  The separate distributions of {\it warm} (24 sources) and {\it
cold} (32 sources) sub-populations are indicated
with the (grey) open and (white) hatched histograms, respectively.
}
\label{f60100}
\end{figure*}

The far-IR luminosity of the IDS/ISOCAM sample
extends  over 3  orders of magnitude (Fig. \ref{eps1}, left panel)  with  a
 mean value, $\log(L_{FIR})=10.2$,  slightly  
lower than the mode of the distribution,  $\log(L_{FIR})\approx 10.5$.
This value is almost the same  as that of the Revised IRAS 60$\mu$m Bright Galaxy 
Sample \cite{San03}, and of a normal spiral galaxy, like the Milky Way
(Mazzei et al. 1992, and references therein).
The ULIRG galaxy, 3-53A,   emits
the maximum far-IR luminosity of 
the sample,  nearly 100 times higher than the median value.
In the same figure (Fig. \ref{eps1}, right panel) we present the rest--frame distribution
of the  $f_{60}/f_{100}$ flux density ratios for our sample. 
This ratio is a measure of the dust temperature which gives information
on the relative fraction of IR light from new and old star populations
\cite{he86, Metal92}. 
 Its mean value,  -0.3,  corresponds to a grain temperature of 
about 36 K, consistent with the value observed for  the bulk 
of  IRAS galaxies \cite{San03, Soi87} .
The $L_{60}/L_{100}$ ratio correlates with  far-IR luminosity, as expected for
the flux density ratio, $f_{60}/f_{100}$ but avoiding any redshift dependence, 
with the most luminous IRAS sources having the largest values of such a ratio 
(see Sanders \&
Mirabel (1996), and references therein).
From the distribution of the
rest--frame luminosity ratio, $L_{60}/L_{100}$ 
we derive a mean  value of -0.08, shown as a (red) continuous line in Fig. \ref{Lfir}.
We divide the sample into two subsamples separated about this mean, 
i.e., a
{\it warm} subsample of 24 IDS/ISOCAM sources with $\log(L_{60}/L_{100})\ge -0.08$, 
and a  {\it cold} subsample of 32 objects having $\log(L_{60}/L_{100}) < -0.08$  \cite{RReC89} .
 Fig. \ref{f60100}   shows 
the far-IR luminosity distributions of
{\it warm} and {\it cold} sources in our sample.
{\it Warm} systems  entail the overall luminosity
range with a mode  (mean) value, 10.45 (10.3), about 3 (1.5) times higher than that of {\it cold} 
sources.

 We use the far-IR luminosity  to quantify 
 the star formation rate (SFR). 
According to Chapman et al. (2000),  $L_{TIR}=k \times SFR$ with $k$ ranging from 
1.5 to 4.2, in units of 10$^9$\,L$_{\odot}$\,M$_{\odot}$\,yr$^{-1}$, for Salpeter's IMF with
upper and lower mass limit 100\,M$_{\odot}$ and  0.1\,M$_{\odot}$ respectively
and  $L_{TIR}=1.7\times L_{60}$. 
 We adopt k=4.2 by comparing the SFR as defined above with the Hopkins et
al. (2003) calibration.
\begin{figure*}
\includegraphics[width=.45\textwidth]{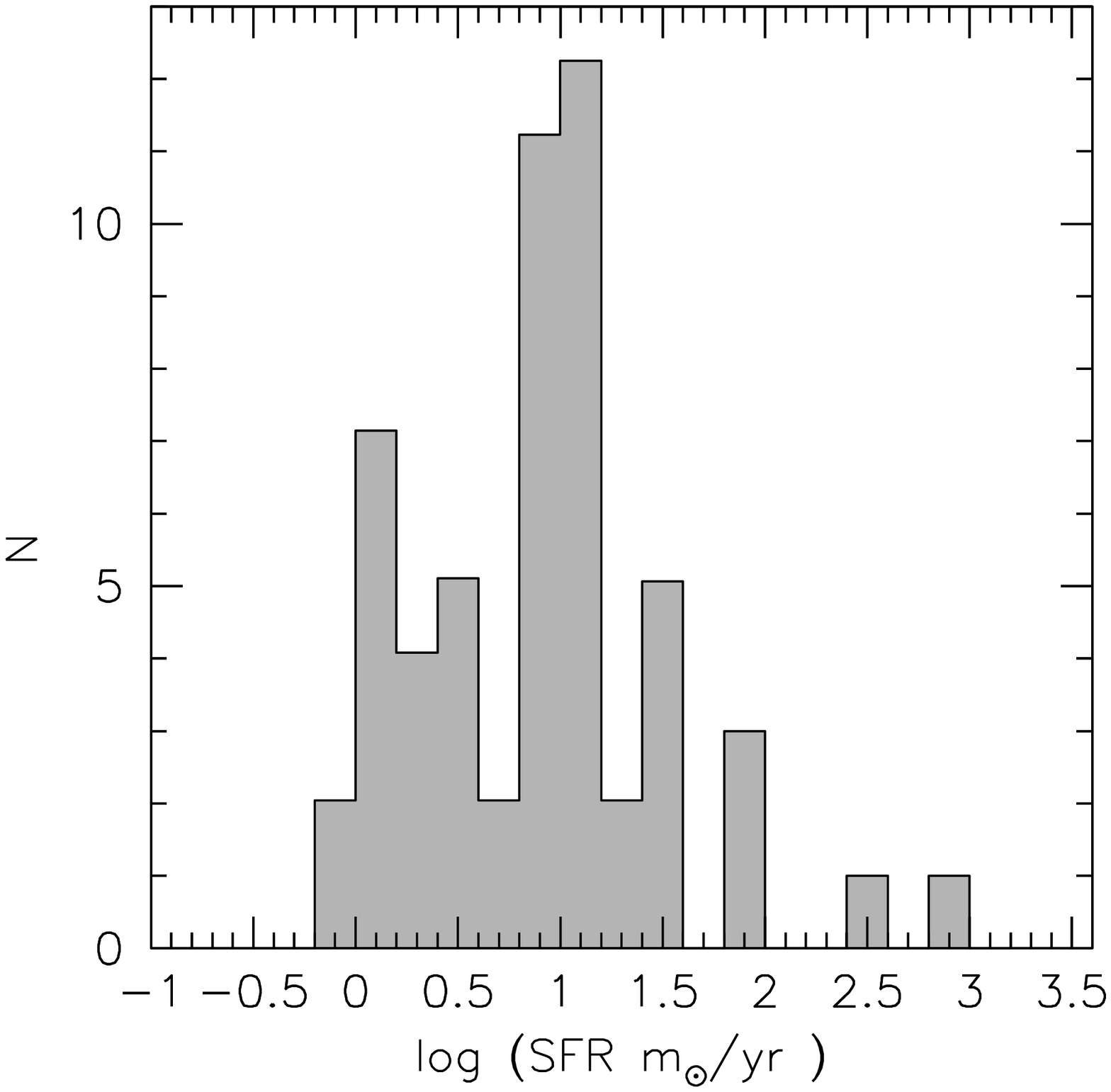}
\includegraphics[width=.45\textwidth]{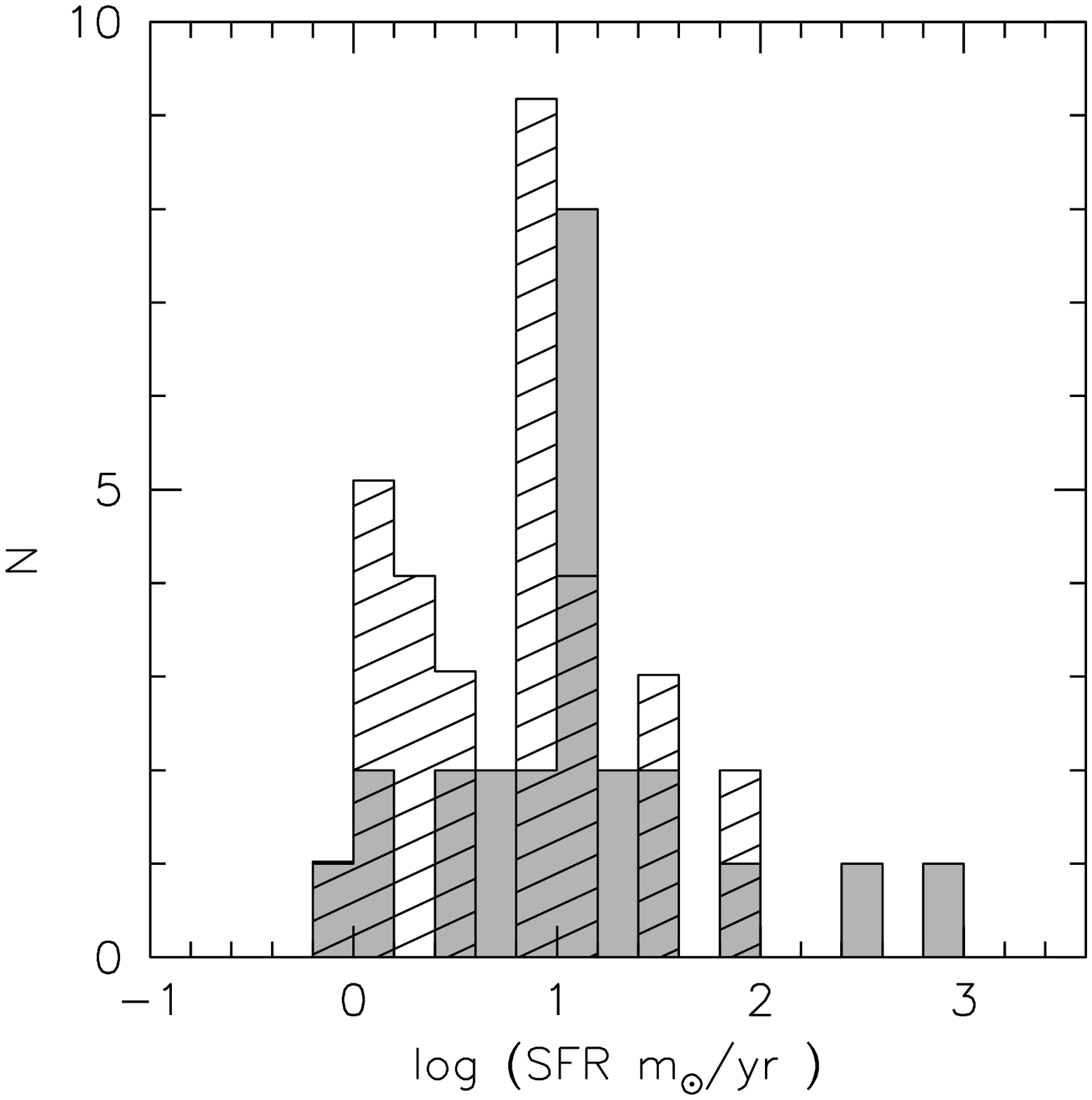}
\caption[]{
{\sl Left:} The SFR distribution for our  sample of 56 sources.
{\sl Right:} The SFR distribution for {\it warm}, 24,  and for {\it cold}, 32 
sources, (grey) open  and (white) hatched histogram respectively.}
\label{eps4}
\end{figure*}
\begin{figure*}
\centering
\includegraphics[width=.45\textwidth]{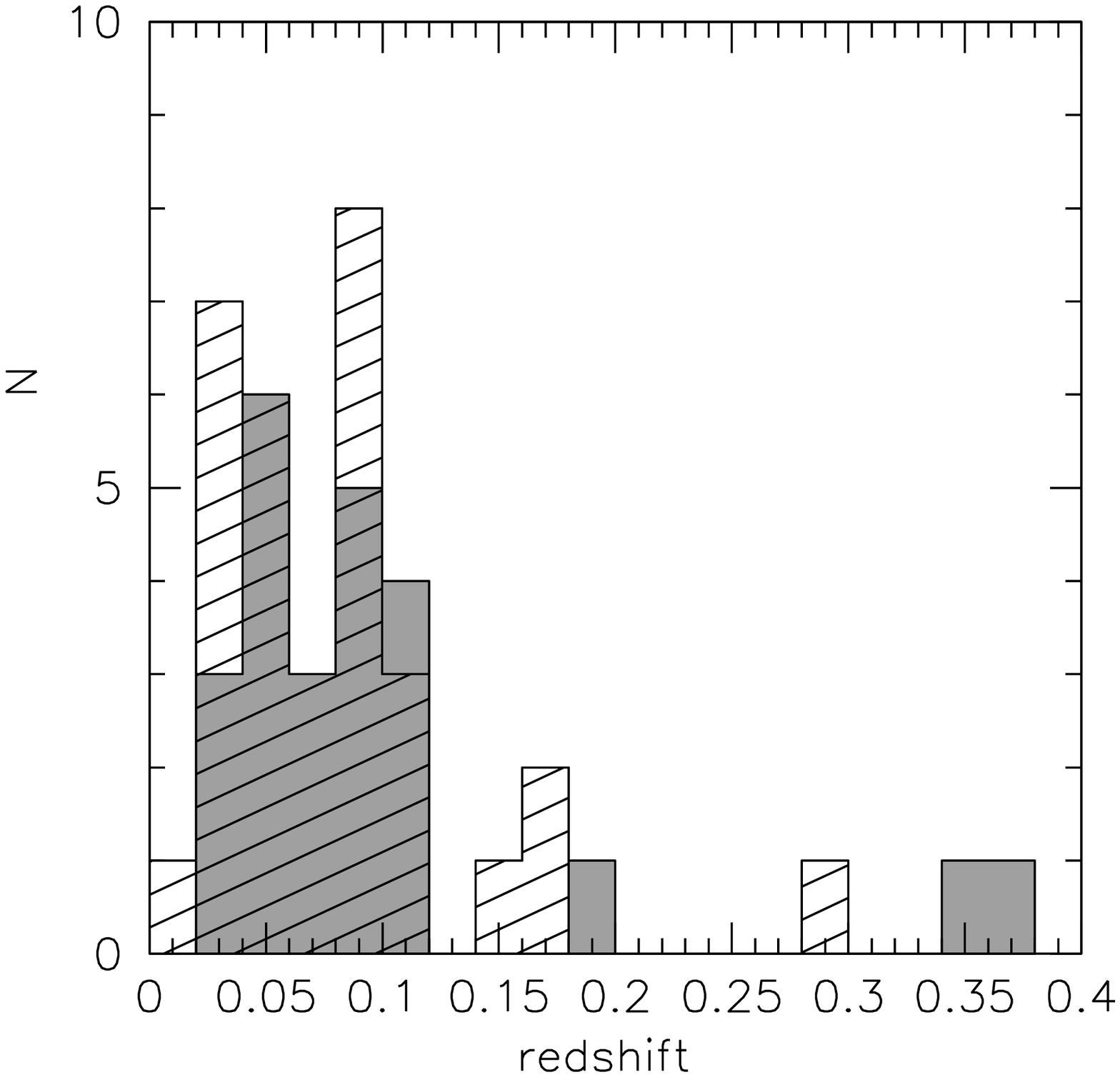}
\caption[]{
The redshift distribution for {\it warm}, 24,  and for {\it cold}, 32 
sources, (grey) open  and (white) hatched histogram respectively.}
\label{eps5}
\end{figure*}
The  more luminous 60\,$\mu$m sources, i.e.,  those with higher SFRs,
correspond to  the more distant systems, 3-96 and
3-53, two {\it warm} sources (Fig. \ref{eps4}). 

 The dust temperatures suggest 
the presence of two galaxy populations: a spiral population
with normal dust temperature (i.e., $<$ 36\,K), mean SFR 
$\simeq 6$\,M$_{\odot}/$yr, mean far-IR luminosity, $\log(L_{FIR})$, $\simeq$ 10.1,
and  mean redshift  0.075, together
with a starburst population characterized by warm dust temperature (i.e., $\ge 36\,$K),  mean SFR
$\simeq 12$\,M$_{\odot}/$yr, mean far-IR luminosity, $\log(L_{FIR}) \simeq 10.3$, and  
mean redshift 0.1 (Fig. \ref{eps5}).

\section{The 60\,$\mu$m luminosity function}\label{lffir}

The observed flux $S_\nu$ is related to the rest-frame 
luminosity by: 

\begin{equation} \label{eq:L}
L_\nu = 4\pi \,\,D_L^2 \,\, S_\nu \,\, / k\left(L,\, z\right)
\end{equation}

\noindent where $D_L$ is the luminosity distance, computed according our
cosmological model (\S \ref{intro}),
and $k\left(L,\,z\right)$ is the \mbox{$K$-correction} defined as:

\begin{equation} \label{eq:k}
k \left(L, \, z \right) = 
\frac{ \left(1+z \right) \,\, L_{\nu (1+z)} } { L_\nu }
\end{equation}
For the most distant galaxies ($z > 0.2$) the correction exceeds  20\% of
the luminosity.

\begin{figure*}
\centering
\includegraphics[width=.45\textwidth]{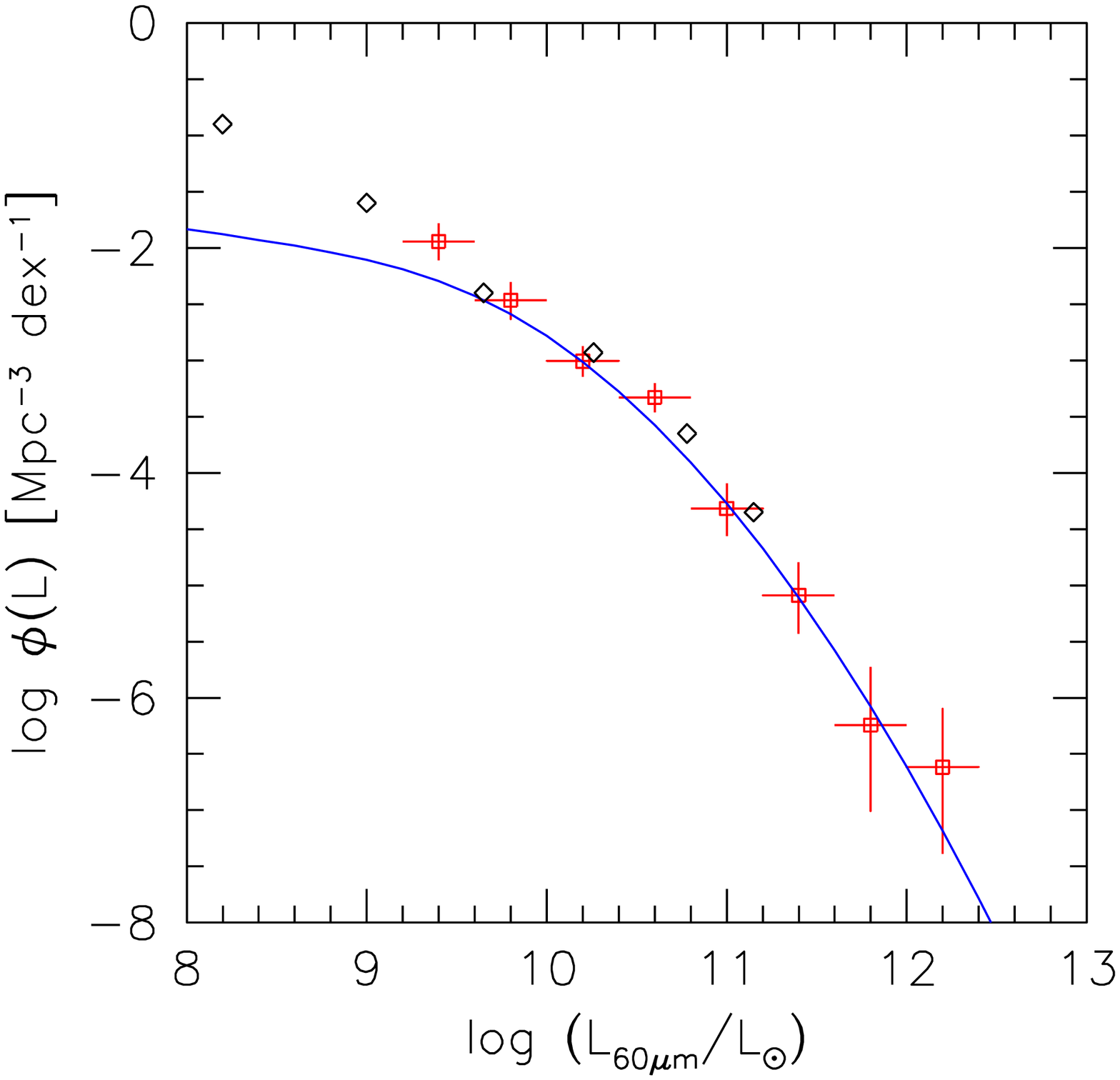}
\includegraphics[width=.45\textwidth]{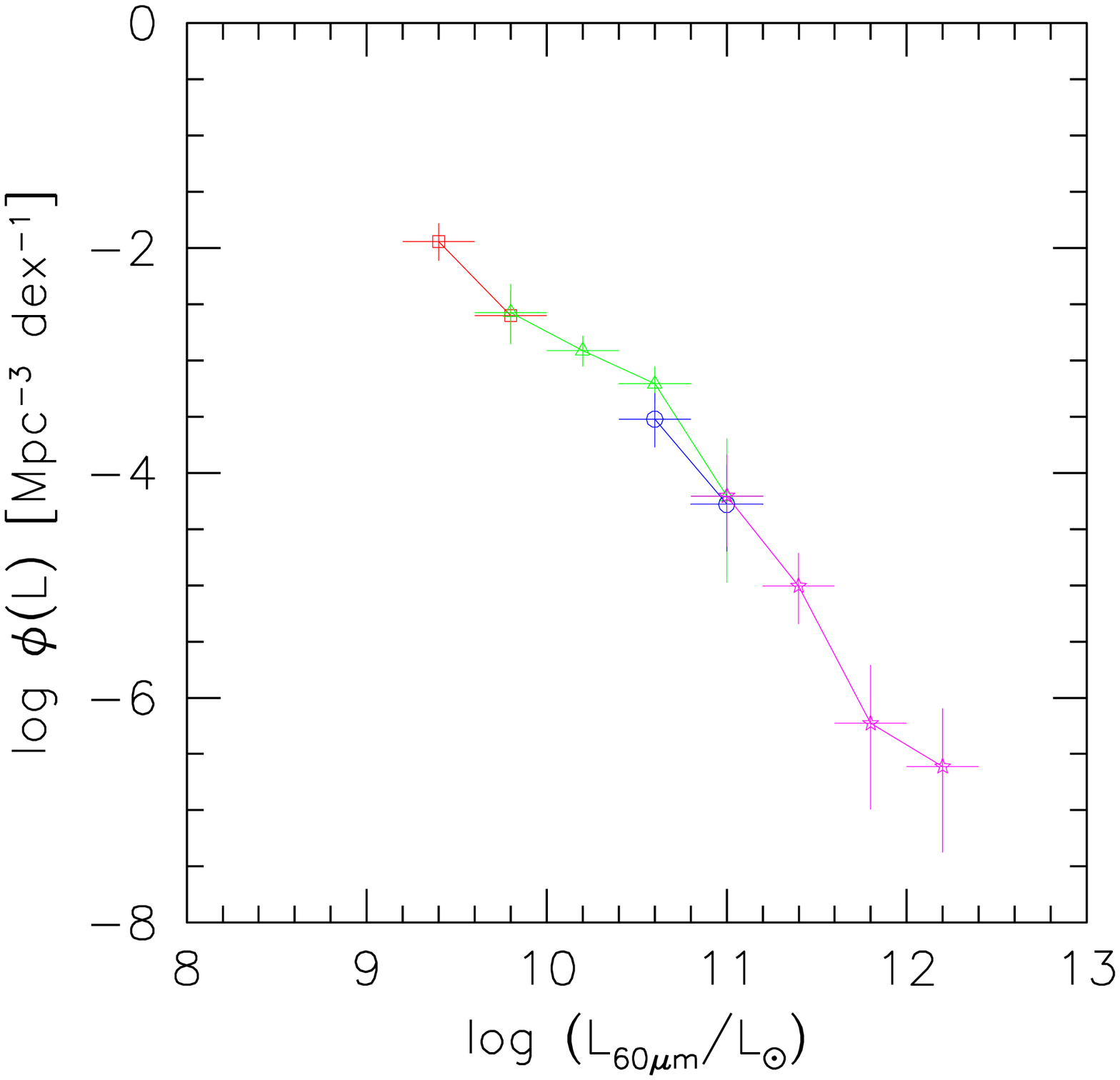}
\caption[]
{{\sl Left:}The rest--frame 60\,$\mu$m  LF for our sample of 56 IDS/ISOCAM sources
(red filled
squares) compared with LF of IRAS PSCz galaxies from  Takeuchi et al. (2003, 2004; 
blue continuous line)
and with LF of Spitzer extragalactic First Look Survey from Frayer et al. (2006) (open diamonds). The bin size is $\Delta\log
L_{60}=0.4$. {\sl Right:}
LF split into different redshift bins: asterisks (red) for $z<0.05$, 
triangles (green) for  $0.05\le z < 0.010$,  open circles (blue) for $0.010\le z < 0.015$, and
stars (magenta) for $z>0.15$.}
\label{lf60}
\end{figure*}

Fig.\ref{lf60}  shows our derived 60$\,\mu$m  differential luminosity
function, i.e., the co-moving number density of sources per logarithmic luminosity
interval,  $\phi(L_{60},z) \Delta$ log$L_{60}$  where $L_{60}=\nu L_{\nu}$,
calculated using the Schmidt-Eales $1/V_{max}$ estimator proposed by Schmidt
(1968) and improved by Felten (1976) and Eales (1993). We assume
 Poisson errors, as tabulated by Gehrels (1986).  $L_{60}$, i.e., the rest--frame luminosity, 
 has been K--corrected as specified in \S\,\ref{farir}. 

According  to the   $V/V_{max}$ test \cite{Schm68}, sources 
drawn from a population uniformly distributed in space should have
a mean value of $V/V_{max}$ equal to 0.5.  Larger (smaller) values of $V/V_{max}$  
 are the result of  a strong increase (decrease) of the co-moving space density 
of sources with redshift. We find $<V/V_{max}>=0.51\pm 0.06$, consistent with a uniform
distribution of sources.
The IDS/ISOCAM sample is ten times 
deeper in flux density than the PSCz catalog  and 100  times deeper than the IRAS 60$\mu$m
Bright Galaxy Sample \cite{San03}
but no signatures of evolution arise from our sample.

The LF we derive  is in good agreement with previous works based 
on  the IRAS PSC \cite{Sau, tsu, tsu04}.
Takeuchi et al. (2003, 2004) 
revised the work by Saunders et al. (1990) by enlarging their galaxy sample 
to 15411 galaxies  from the PSCz \cite{Sau00}
with a flux limit of 600 mJy and a redshift range between 0 and 0.07.
 Their analytic fit, shown as a continuous line in Fig. \ref{lf60}, is based
on the same parameterization as Saunders et al. (1990):
\begin{equation} \label{eq:saunders}
\phi(L_\nu) = \phi_* \left(\frac{L_\nu}{L_*}\right)^{1-\alpha}
\exp \left[ -\frac{1}{2\sigma^2} \log_{10}^2 \left(1+\frac{L_\nu}{L_*}\right)\right]
\end{equation}
with parameters: $\alpha=1.23\pm 0.04$, $L_{*}=(4.34\pm 0.86)\times 10^8 h^{-2}$\,L$_{\odot}$,
$\sigma=0.724\pm 0.01$, and $\phi_*=(2.60\pm 0.30)\times 10^{-2}h^3\,$Mpc$^{-3}$
\cite{tsu, tsu04}.
Our results extend
over  three orders of magnitude in luminosity, from  $\log(L_{60}/L_{\odot}) \simeq$ 9
up to more than 12.
 In the range  where the samples overlap our findings  agree 
with the recent determination by Frayer et al. (2006, diamonds in Fig.
\ref{lf60}), based on a complete 
sample of 58 sources with $S_{70}>50$\,mJy and $z<0.3$, drawn from the {\it Spitzer extragalactic First Look
Survey}. \\
Despite the fact that the redshift range  exceeds $z=0.3$, our LF does not show any evidence of
evolution.\\
A more direct test for evolution has been performed by computing LF in four
different redshift bins assuming no evolution  (see  Fig. \ref{lf60}, right
panel,  and Table \ref{vvmax}).  The results   are fully consistent
with this assumption.
 
\begin{table*}
\caption{$60\,\mu$m luminosity function: V/V$_{max}$ in different redshift bins.
\label{tbl-vmax}}
\label{vvmax}
\begin{center}
\begin{tabular}{rrrrrrr}
\hline\hline
redshift bin & V/V$_{max}$  \\
\hline
0.00$\le z <$ 0.05 & 0.490$\pm$ 0.04\\
0.05$\le z <$ 0.10 & 0.471$\pm$ 0.08\\
0.10$\le z <$ 0.15 & 0.444$\pm$ 0.07\\
$z>$ 0.15 & 0.315$\pm$ 0.08\\
\hline
\end{tabular}
\end{center}
\end{table*}

\section {The 15$\mu$m  luminosity function}

Deep surveys at 15$\mu$m carried out using  ISO  
\citep[i.e.,][]{Elbazetal99, Flores99, Larietal01, Metcalfe2003} 
seem to require strong
evolution of 15$\mu$m sources starting from redshift 0.5
(see Lagache et al. (2005) and references therein). Several
evolutionary models were developed to explain these results
\citep[e.g.,][]{Fetal01,RR01}
trying to fit IR/submillimiter source counts with different
degrees of success. Nevertheless none of them is based on a local LF obtained 
from 15$\mu$m data, since the only available data until few years ago came from
IRAS 12$\mu$m photometry \citep{Rush93, xu98, Fang98}. 

A first attempt to build up the 15$\mu$m LF of a NEPR subsample was made by
Xu (2000) using the bi-variate method
to translate the 60$\mu$m local LF of IRAS galaxies \citep[]{Sau}
to 15$\mu$m. Its sample comprises
64 sources  detected both at 60$\mu$m \cite{HH87} and at 15$\mu$m \cite{Aus99}, 
with redshifts measured by Ashby et al. (1996). Xu fitted the result obtained
(see its Table~2 and Fig.~3), with a pure luminosity 
evolution model $L \propto (1+z)^{4.5}$. Nevertheless he said 
that it must be considered as a preliminary work as discussed in
\S\,\ref{intro}.

A recent determination of the 15$\mu$m local LF  was made by 
Pozzi et al. (2004)  using the available data on the 
southern  fields, S1 and S2, of the ELAIS survey \cite{Oliver00}.
Their data sample entails 150 galaxies with redshift $z \leq 0.4$, excluding 
sources classified as AGNs (both type one and type two). The 15$\mu$m LF was
calculated with a parametric maximum likelihood method.
Pozzi et al. (2004) separate  spirals from starbursts
using optical/mid-IR ratios, assuming that  Starbursts are the more
mid-IR luminous galaxies  with, on average, larger  
$\log(L_{15}/L_{R}$) ratios. 
They  estimate the 15$\mu$m local LF in the range
 $ 7.8 \le \log (L/L_\odot) \le 10.6$. Their findings are that 
the 15$\mu$m LF of   spirals  is consistent
with no evolution, ($<V/V_\mathrm{max}> = 0.55 \pm 0.03$), but that the
value $<V/V_\mathrm{max}>= 0.64 \pm 0.03$ measured for the 
starbursts suggests that this population is in fact evolving.

\subsection {The bi-variate method}

Since our sample is flux limited at 60$\mu$m, but not at 15$\mu$m, we used the 
bi-variate method to calculate the 15$\mu$m LF of our sample. 
It was
obtained by a convolution of the 60$\mu$m LF with the log($L_{15}/L_{60}$) distribution:
\begin{equation}
\phi(\log L_{15})= \int \phi(\log L_{60})\, P(\log (L_{15}/L_{60}))\, d (\log L_{60})
\end{equation}
\noindent 
where $P(log(L_{15}/L_{60}))$ is the conditional probability
function that gives the distribution of log($L_{15}$) around the mean 
15$\mu$m luminosity $\langle log(L_{15}) \rangle$ at a given 60$\mu$m luminosity
log($L_{60}$).

The distribution per unit interval of the logarithm of the luminosity ratio
is well described by Type~I Pearson's curves \citep{Pearson1924, Elderton1969}:
\begin{equation}
y = y_0 \,\, (1+x/a_1)^{m1}\,\, (1-x/a_2)^{m2}
\end{equation}
\noindent
where $x = \log(L_{15}/L_{60})$, with $-a_1 \leq x \leq a_2$,
and origin at the position of the peak of the distribution ({\it mode}). 
The values of the parameters are given in Table \ref{biva}, together with the 
mean, 
the standard deviation $\sigma$, the skewness ($\mu_3^2/\mu_2^3$) and the
kurtosis ($\mu_4/\mu_2^2-3$) of the distribution 
($\mu_i$ is the $i$-th moment about the
mean). The quality of the fit is quantified by the value of $\chi^2$ per degree
of freedom ($\chi_\nu$), given in the last column and computed adopting 
the Levenberg-Marquardt method as implemented in Press et al. (1992); $\sigma_p$
are the errors on the Pearson's parameters.

To perform these analyses we remove from our sample the  sources
3-70 and 3-96 with AGN properties (see \S$\,$\ref{farir})
and used the parametric solution of the 60$\mu$m LF derived by Takeuchi et al. 
(2003, 2004), which  agrees well  with our results (see  \S$\,$\ref{lffir},
eq. \ref{eq:saunders}). 

\begin{table*}
\caption{Coefficients and parameters describing the distribution function, 
Pearson~I curve of our sample.}
\label{biva}
\begin{center}
\begin{tabular}{llllllllllll}
\hline
\hline
Case        & $y_0$ & $a_1$ & $a_2$ & $m_1$ & $m_2$ & Mode  & Mean  & $\sigma$ & Skew & Kurt & $\chi_\nu^2$ \\
\hline
All         & 82.8  & 12.9  & 0.43  & 83.3 & 2.72  & -0.60 & -0.69 & 0.24   & 0.52 & 0.52 & 0.77 \\
$\sigma_p$   & 16.5  & 1.56  & 0.04  & 10.2 & 0.39  &  0.04 &  &   &  &  &  \\
\hline
\end{tabular}
\end{center}
\end{table*}
\begin{table*}
\caption{Parameters describing the LF of our sample.}
\label{bivaLF}
\begin{center}
\begin{tabular}{llllllllllll}
\hline
\hline
Case       & $\phi_{*}/10^{-3}\,Mpc^{-3}$    & $L_{*}/10{^8}\,L_{\odot}$    &$\phantom{)))))} \alpha$ & 
$\phantom{)))))} \sigma$\\ 
\hline
All  & $5.32\pm .40$  & $2.00\pm .18$ & $1.241\pm .021$ & $0.748\pm .010$  \\
\hline
\end{tabular}
\end{center}
\end{table*}

\begin{figure*}
\centering
\includegraphics[totalheight=0.40\textheight]{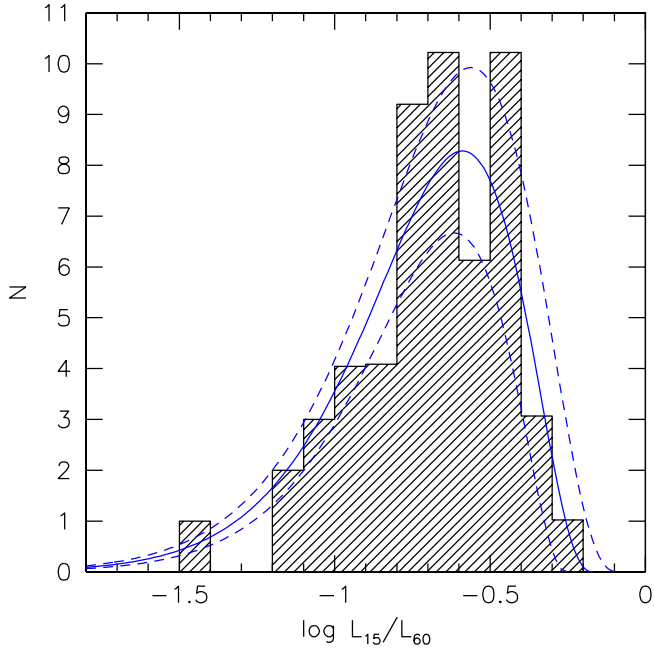}
\vspace{-0.75cm}
\caption{\small Distribution of the $\log L_{15}/L_{60}$ ratio for 54 IDS/ISOCAM
sources (see text). The
bin size is $\Delta(\log L_{15}/L_{60})=0.1$. The
fitting function is a Type~I Pearson  whose parameters, optimized following the
 Levenberg-Marquardt method (see text),
are reported in Table \ref{biva}; dashed lines show the same 
curves with 1$\sigma$ error in the parameters.}
\label{pearson}
\end{figure*}
\begin{figure*}
\centering
\includegraphics[width=.5\textwidth]{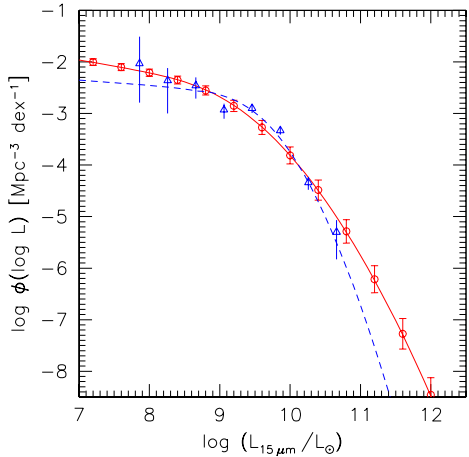}
\caption[]{ The 15$\,\mu$m differential LF  (red continuous line) for the
IDS/ISOCAM sample (54 sources) compared with 
that of Pozzi et al. (2004) (blue dashed line and open
triangles). The error bars  correspond 
to one $\sigma$ error  of the Pearson's curve parameters.
}
\label{eps3}
\end{figure*}

Table \ref{bivaLF} lists the parameters  defining  our 15\,$\mu$m local galaxy LF
using the same parametric form as in eq. (\ref{eq:saunders}) suggested by
 Pozzi et al. (2004).
The Pozzi LF extends from 7.8  up to  10.6 in log(L$_{15}/$L$_{\odot})$, whereas  
our convolution  extends over more than six  orders of 
magnitude  and  100 times deeper  (Fig. \ref{eps3}).
Our results agree with those by Pozzi et al. (2004) 
up to their luminosity limit (10.6), however, beyond such a luminosity
the two parametric LFs diverge so that at  log(L$_{15}/$L$_{\odot}$)=11  we
expect  10 times more sources than Pozzi et al. (2004).

\subsection{Spiral and starburst populations}\label{sbssect}
\begin{figure*}
\centering
\includegraphics[totalheight=0.27\textheight]{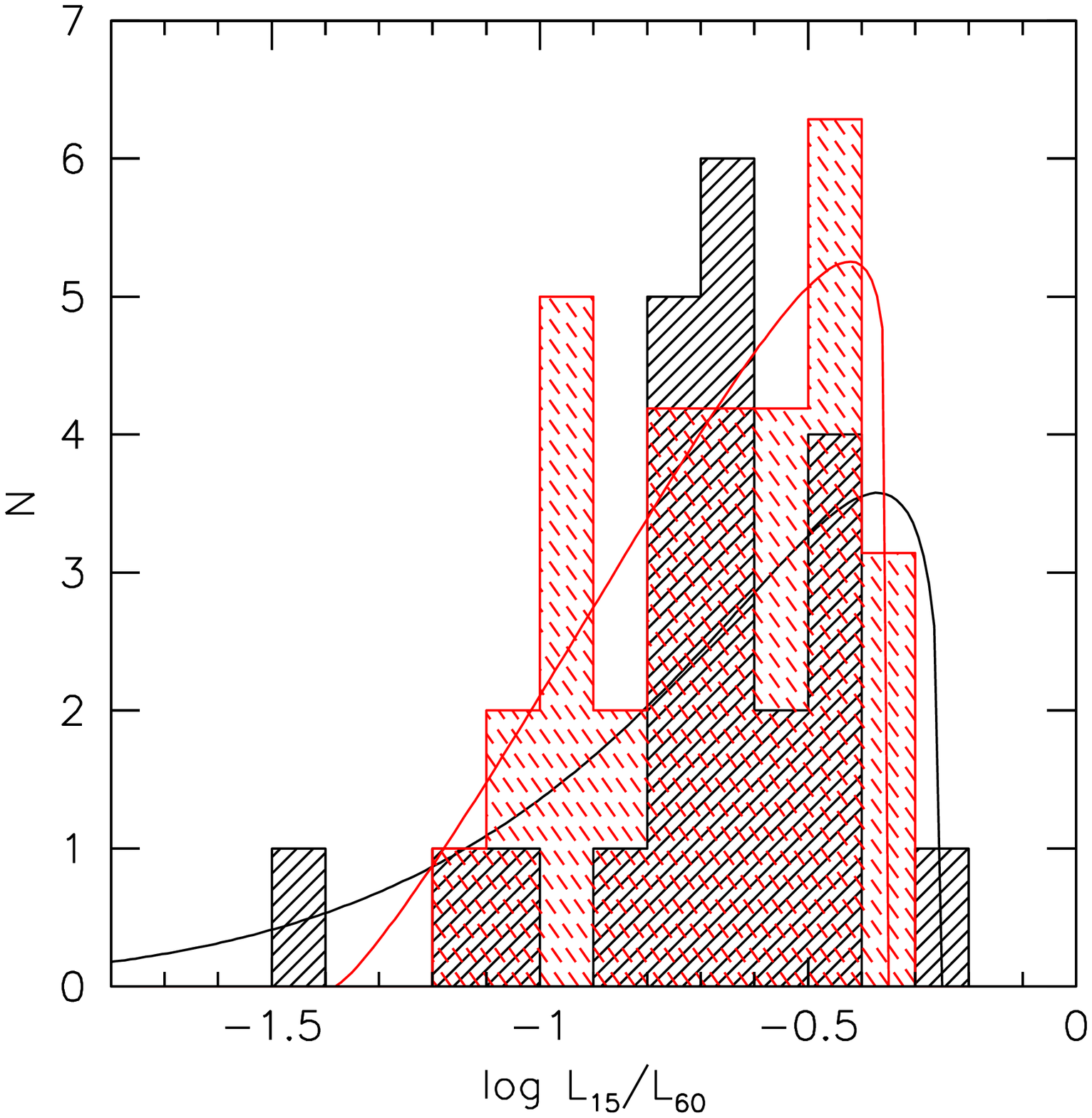}
\includegraphics[totalheight=0.27 \textheight]{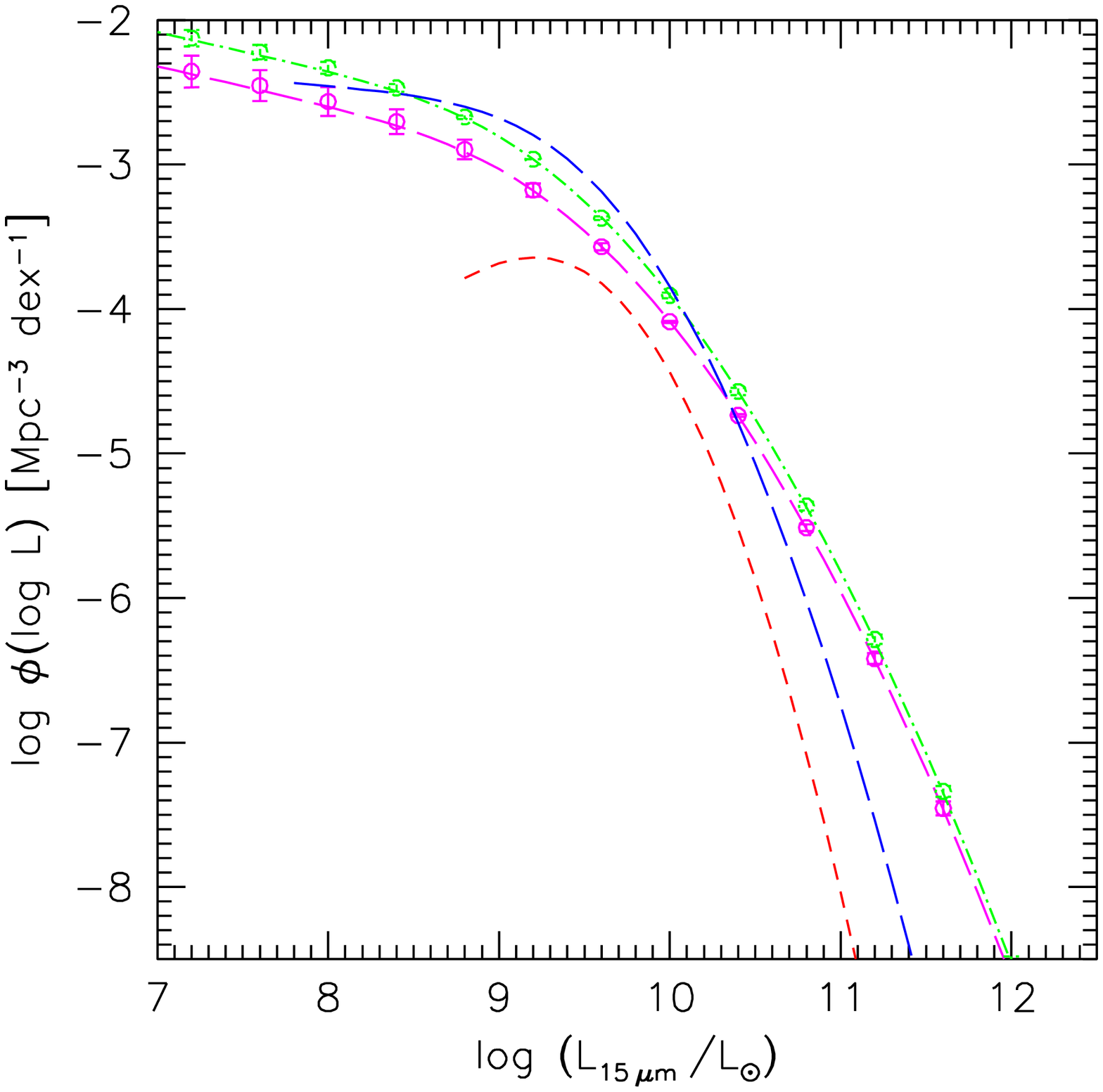}
\vspace{-0.75cm}
\caption{
{\sl Left:} Distribution of the $\log L_{15}/L_{60}$ ratios for starbursts 
(black continuous line), 22 sources, and  spirals (red dashed line) 32, sources, 
selected on the basis of their far-IR colors. The
bin size is $\Delta(\log L_{15}/L_{60})=0.1$ and the parameters of the 
fitting functions, Type~I Pearson curves, are in Table \ref{bivasb} together 
with one $\sigma$ errors. 
{\sl Right:} The contribution of such galaxy populations 
to the rest-frame 15$\mu$m LF (see Table \ref{bivas}): 
starbursts (magenta long--dashed dotted line) and spirals (green dot-dashed
line).  Results are compared with those by
Pozzi et al. (2004):
starbursts (red short--dashed line) and spirals (blue long-dashed line).
} 
\label{pearson_sbs}
\end{figure*}

Our sample includes 22 galaxies with rest-frame  {\it warm}  $L_{60}/$L$_{100}$
ratios  and 32 galaxies with {\it cold} ratios 
 after excluding AGN (see \S$\,$\ref{farir}). 
They define
two different populations,  starburst and spiral galaxies respectively,
as far as dust properties are concerned.
Their distributions per unit interval of the 
logarithm  of the luminosity ratios $\log(L_{15}/L_{60}$),
shown in Fig. \ref{pearson_sbs} (left panel), are fitted well by  Pearson's curves
whose parameters are in Table \ref{bivasb}. 
Their  15\,$\mu$m LF fits, i.e., eq. (\ref{eq:saunders}), 
are in Fig. \ref{pearson_sbs} (right panel) and  the LF parameters  are given in
Table \ref{bivas}.
We find that   both  populations 
contribute to the  faint end of  the LF.
Spiral galaxies overcome starbursts  by less than a factor of
two.
Such a factor is slightly reducing with  luminosity, from 1.8 at 
$\log(L_{15}/L_{\odot})=8$
to 1.3 at $\log(L_{15}/L_{\odot}=11.8)$.
Our results differ from those of Pozzi et al.(2004),  in particular predictions
concerning starburst population.  However
they  used  optical/mid-IR ratios
instead of far-IR ratios to disentangle starbursts and spirals,
assuming that starbursts  are the more
mid-IR luminous galaxies  with, on average, larger  
$\log(L_{15}/L_{R}$) ratios.
\begin{table*}
\caption{
Coefficients and parameters describing the distribution function, 
Pearson~I curve, of starburst and spiral populations.}
\label{bivasb}
\begin{center}
\begin{tabular}{llllllllllll}
\hline
\hline
Case        & $y_0$ & $a_1$ & $a_2$ & $m_1$ & $m_2$ & Mode  & Mean  & $\sigma$ & Skew & Kurt & $\chi_\nu^2$ \\
\hline
Starbursts & 35.7  & 4.30  & 0.14  & 9.11 & 0.25  & -0.39 & -0.70 & 0.25   & 1.99 & 2.39 & 1.70 \\
$\sigma_p$   & 12.3  & 1.55  & 0.02  & 3.56 & 0.18  &  0.002 &  &   &  &  &  \\
Spirals   & 43.2  & 1.47  & 0.32 & 4.43 & 0.19  & -0.64 & -0.64 & 0.23   & 0.05& -0.84 & 0.77 \\
$\sigma_p$   &11.6  & 0.58  & 0.01  & 1.89 & 0.11  &  0.01 &  &   &  &  &  \\
\hline
\end{tabular}
\end{center}
\end{table*}

\begin{table*}
\caption{
Parameters describing the LF of starburst and spiral
populations.}
\label{bivas}
\begin{center}
\begin{tabular}{lllll}
\hline
\hline
Case       & $\phi_{*}/10^{-3}\,Mpc^{-3}$   & $L_{*}/10{^8}\,L_{\odot}$   & $\phantom{)))))} \alpha$ & 
$\phantom{)))))} \sigma$\\ 
\hline
Starbursts  & $1.89\pm .03$  & $2.96\pm.03$ & $1.275\pm .001$ & $0.740\pm .015$  \\
Spirals     & $3.36\pm .07$  & $2.21\pm.07$ & $1.236\pm .006$ & $0.739\pm .027$  \\
\hline
\end{tabular}
\end{center}
\end{table*}

\begin{figure*}
\centering
\includegraphics[totalheight=0.35 \textheight]{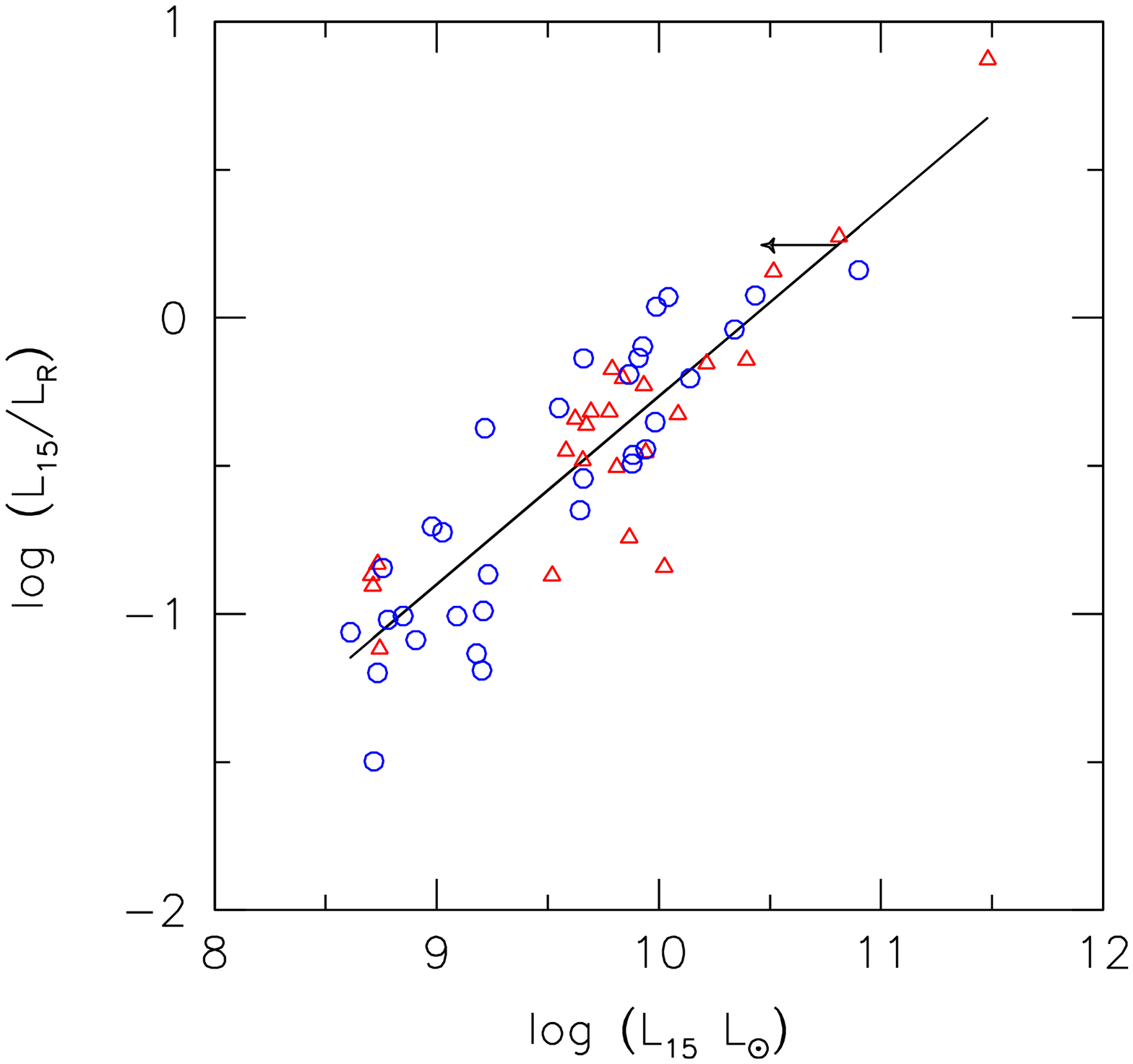}
\caption{\small The behavior of the $\log L_{15}/L_{R}$ ratio for {\it warm}  sources (24, red
triangles), and for {\it cold} sources (32, blue open circles) of
IDS/ISOCAM sample;
the continuous line corresponds to the least--square fit (see eq.
\ref{leastsq}).}
\label{L15LR}
\end{figure*}
To  investigate further  this point, we repeated our analysis using the same
criterion of Pozzi et al.  to discriminate  starbursts from spirals. 
We used  APS catalog  (http://aps.umn.edu) R-band magnitudes for the 37.5\% 
 of our sample 
galaxies that lacked them (see \S$\,$\ref{intro}), applying a correction   of 
-0.75 magnitudes to bring the  APS zeropoint into agreement with our own common sources.

\begin{figure*}
\centering
\includegraphics[totalheight=0.27\textheight]{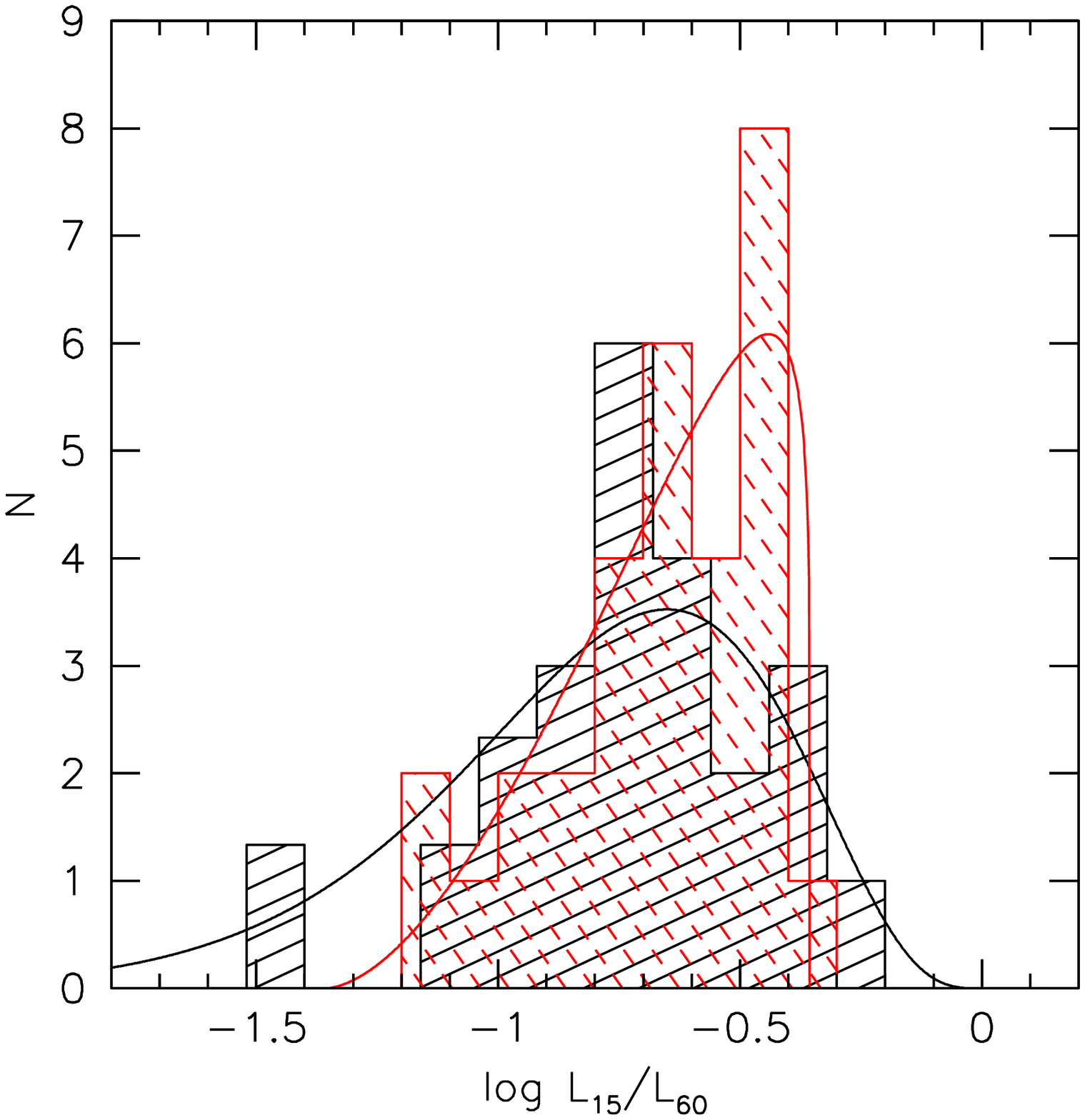}
\includegraphics[totalheight=0.27\textheight]{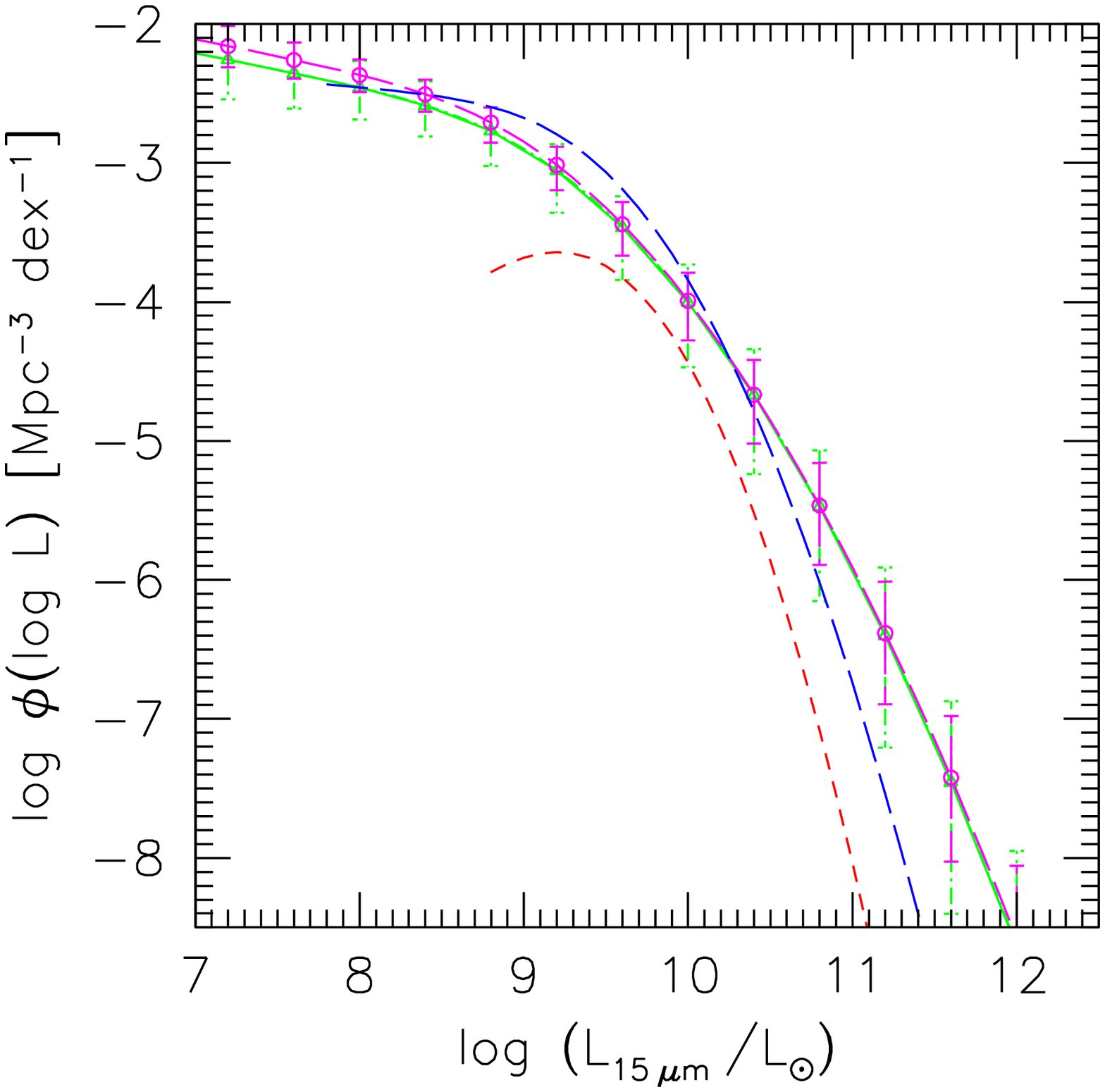}
\vspace{-0.75cm}
\caption{\small 
{\sl Left:} The distribution of the $\log L_{15}/L_{60}$ ratio for starbursts 
(black continuous line), 24 sources, and for spirals (red dashed line) 30 sources,
selected with the same criterion as Pozzi et al. (2004). The
bin sizes are $\Delta(\log L_{15}/L_{60})=0.12$ and 0.1 respectively. 
The parameters of the 
fitting functions, Type~I Pearson curves, are in Table  \ref{bivasb_new}.
{\sl Right:} The contribution of such galaxy populations to the rest-frame 
15$\mu$m LF (see Table \ref{bivas_new}): 
starbursts (magenta long--dashed dotted line) and spirals (green dot--dashed
line). Our results are compared with those by
Pozzi et al. (2004):
starbursts (red short--dashed line) and spirals (blue long-dashed line).
}
\label{lf15_sbs}
\end{figure*}

Fig. \ref{L15LR} shows the rest--frame   $L_{15}/L_{R}$
ratios vs. $L_{15}$ luminosity. From a least--square--fitting procedure 
we find:
\begin{equation}
\centerline { $L_{15}/L_{R}=(0.64 \pm 0.04)\times L_{15} -(6.67 \pm 0.43)$}
\label{leastsq}
\end{equation}
with a dispersion of 0.20\,dex.  Our fit  is 
steeper, 0.64 instead of 0.5, than that of  Pozzi et al. (2004). 
Following Pozzi et al. (2004), we assume $L_{15}/L_{R}=-0.4$ as the nominal 
separation between spiral and starburst populations. 
{\it Warm} and {\it cold} galaxies, selected on the basis of their far--IR colors, 
are mixed in Fig. \ref{L15LR}.  Such a criterion selects
different galaxies in both the populations.
 Now there are 30 spirals and 24 starbursts 
in our sample. The distributions per unit interval of the logarithm 
of their luminosity ratios $\log(L_{15}/L_{60}$), are fitted again by Type~I 
Pearson's curves (Fig. \ref{lf15_sbs}, left panel)
whose parameters are in Table 
\ref{bivasb_new}. Table  \ref{bivas_new} shows those of their 
15\,$\mu$m LF fits (Fig.   \ref{lf15_sbs}, right panel).

\begin{table*}
\caption{
Coefficients and parameters describing the distribution functions, 
Pearson~I curves,  of starburst and spiral populations.}
\label{bivasb_new}
\begin{center}
\begin{tabular}{llllllllllll}
\hline
\hline
Case        & $y_0$ & $a_1$ & $a_2$ & $m_1$ & $m_2$ & Mode  & Mean  & $\sigma$ & Skew & Kurt & $\chi_\nu^2$ \\
\hline
Starbursts & 35.11  & 11.5  & 0.62  & 61.2 & 3.45  & -0.63 & -0.74 & 0.29   & 0.50 & 0.56 & 0.86 \\
$\sigma_p$ & 10.7  & 2.29  & 0.12  & 12.3 & 0.74  &  0.11 &  &   &  &  &  \\
Spirals    & 60.3  & 10.0 & 0.33  & 47.8 & 0.94  & -0.60 & -0.67 & 0.21   & 0.41& -0.34 & 1.03 \\
$\sigma_p$ & 13.0  & 2.61  & 0.13  & 12.6 & 0.47  &  0.11 &  &   &  &  &  \\
\hline
\end{tabular}
\end{center}
\end{table*}

\begin{table*}
\caption{
Parameters describing the LF of starburst and spiral populations.}
\label{bivas_new}
\begin{center}
\begin{tabular}{lllll}
\hline
\hline
Case       & $\phi_{*}/10^{-3}\,Mpc^{-3}$   & $L_{*}/10{^8}\,L_{\odot}$   & $\phantom{)))))} \alpha$ & 
$\phantom{)))))} \sigma$\\ 
\hline
Starbursts  & $3.82\pm .62$  & $1.81\pm.45$ & $1.245\pm .048$ & $0.761\pm .028$  \\
Spirals     & $2.92\pm .88$  & $2.29\pm.85$ & $1.234\pm. 084$ & $0.742\pm .037$  \\
\hline
\end{tabular}
\end{center}
\end{table*}

Both such populations 
contribute to the  faint end of  the LF.
 The space densities of starbursts are just slightly greater than those of spiral 
 galaxies.         

\section{Conclusions}
Combining our observations \cite{DV05} with those  by Ashby et al. (1996)
we have reliable identifications and spectroscopic redshifts for 100\% of
the complete  far--IR selected subsample comprising 56 IDS sources with $S_{60}>80$mJy
\cite{Maetal01}.
The  redshift distribution shows a tail extending up to $z\simeq 0.37$, in
particular  $\simeq$ 26\% of the sources have redshifts  $z>0.1$.


 To fully exploit the potential of this sample,  
  ten times deeper than the IRAS PSC, thus less liable
to the effect of local density inhomogeneity, for investigating galaxy
evolution, we calculate the 60$\mu$m LF 
using the $1/V_{max}$ method. 
Current estimates are based on rather shallow
samples. Even though our  sample is five times deeper 
in redshift than the PSCz  \cite{Sau00} used by
Takeuchi et al. (2003), our LF agree with their
determination \cite{tsu, tsu04}
and with that by Frayer et al. (2006) based on a complete 
sample of galaxies with $S_{70} > 50$~mJy drawn from the Spitzer
{\it extragalactic First Look Survey}.
Despite the fact that our redshift range exceeds $z=0.3$, and
our 60$\mu$m LF   extends up to $\log L_{60}=12$, whereas  that by Frayer et al.(2006) up to 
$\log L_{60}\simeq 11$,  it does not show any evidence of  evolution.
The $V/V_{max}$ test gives a value consistent with an uniform
distribution ($V/V_{max} = 0.51 \pm 0.06$). Moreover,
a more direct test for evolution has been performed by splitting the LF in 
different redshift bins assuming no evolution.  The results   are fully consistent
with this assumption.

We present the bi-variate 15$\mu$m LF,  one of the few determinations based on
ISO data, by convolving the 60$\mu$m LF with the luminosity
ratio distribution, $L_{15}/L_{60}$ of our sample.
 This extends to luminosity 100 times higher than before.
Our result agrees with the recent determination by Pozzi et al. (2004) in the 
common range of luminosity, i.e., from 7.8  up to  10.6 in 
$\log(L_{15}/L_{\odot})$. 
However, above $\log(L_{15}/L_{\odot})=10.6$, 
the two parametric LFs diverge so that at  $\log(L_{15}/L_{\odot})=11$  we
expect  10 times more sources than Pozzi et al. (2004).

In order to investigate  the role of galaxy populations on such a result,
we disentangle starbursts and  spirals on the basis both
of their far-IR dust temperature, and of their $L_{15}/L_{R}$ ratios,
as assumed by Pozzi et al. (2004).
Such criteria select galaxies with different dust
properties, as we point out in \S$\,$\ref{sbssect}.
Nevertheless,
we find that both galaxy populations contribute to the
faint end of the rest-frame 15$\mu$m LF, even though we adopt the
same criterion as Pozzi et al. (2004). Moreover, in this case,
 above $\log(L_{15}/L_{\odot})=9$ our findings are that starbursts and spirals
give almost the same contribution
to the mid--IR  LF whereas Pozzi et al. (2004) predict that  spirals
contribute 5-10 times more than  starbursts.

\begin{acknowledgements}
We thank the referee, M. Ashby, for an unusually helpful and
quick report,  Gianfranco De~Zotti and Herv\'e Aussel for useful discussions.
Work supported in part by MIUR (Ministero Italiano dell'Universit\'a e della
Ricerca) and ASI (Ageanzia Spaziale Italiana).
\end{acknowledgements}




\end{document}